\documentclass[12pt,openright]{article}
\usepackage{graphicx}
\usepackage{amssymb}		
\usepackage{lmodern}
\usepackage[]{algorithmicx}
\usepackage{algpseudocode}
\usepackage{sectsty}
\usepackage{setspace}
\usepackage{lineno}
\usepackage{amsmath}
\usepackage{graphicx}
\def\infinity{\rotatebox{90}{8}}
\sectionfont{\fontsize{13}{15}\selectfont}
\subsectionfont{\fontsize{11}{15}\selectfont}
\usepackage{geometry}
\begin{document}
\title{\textbf{\large Adjacencies in Permutations}
\date{\vspace{-5ex}}
\author{{\footnotesize \footnote{$~~ Corr.~ author$} $^{1,2}$Bhadrachalam Chitturi  $^1$Krishnaveni K S}}}
\maketitle
\begin{center}
{\footnotesize $^1$Department of Computer Science, Amrita Vishwa Vidyapeetham, Amritapuri Campus, Kollam, Kerala 690525, India.}\\
{\footnotesize $^2$Department of Computer Science, University of Texas at Dallas, Richardson, Texas 75083, USA.}
\end{center}

{\footnotesize\textbf{Abstract:}
 A permutation on an alphabet $ \Sigma $, is a sequence where every element in $ \Sigma $ occurs precisely once. Given a permutation $ \pi $= ($\pi_{1} $, $ \pi_{2} $, $ \pi_{3} $,....., $ \pi_{n} $) over the alphabet $ \Sigma $ =$\{ $0, 1, . . . , n$-$1 $\}$ the elements in two consecutive positions in $ \pi $ e.g. $ \pi_{i} $ and $ \pi_{i+1} $ are said to form an \emph{adjacency} if $ \pi_{i+1} $ =$ \pi_{i} $+1. 
The concept of adjacencies is widely used in computation. The set of permutations over $ \Sigma $ forms a symmetric group, that we call P$ _{n} $. The identity permutation,  I$ _{n}$ $\in$ P$_{n}$ where  I$_{n}$ =(0,1,2,...,n$-$1) has exactly n$ - $1 adjacencies. Likewise, the reverse order permutation R$_{n} (\in P_{n})$=(n$-$1, n$-$2, n$-$3, n$-$4, ...,0) has no adjacencies. We denote the set of permutations in P$_{n} $ with exactly k adjacencies with P$_{n} $(k). We study variations of adjacency.
A transposition exchanges adjacent sublists; when one of the sublists is restricted to be a prefix (suffix) then one obtains a prefix (suffix) transposition.
We call the operations: transpositions, prefix transpositions and suffix transpositions as block-moves. A particular type of adjacency and a particular  block-move are closely related.
In this article we compute the cardinalities of P$_{n}$(k) i.e. $ \forall_k \mid $P$ _{n} $ (k) $ \mid $ for each type of adjacency in $O(n^2)$ time.
Given a particular adjacency and the corresponding  block-move, we show that $\forall_{k} \mid P_{n}(k)\mid$ and the expected number of moves to sort a permutation in P$_{n} $ are closely related. Consequently, we propose a  model to estimate the expected number of moves to sort a permutation in P$_{n} $ with a  block-move. We show the results for prefix transposition. Due to symmetry, these results are also applicable to suffix transposition.
}

\section{Introduction}
Sets and multisets are collections of objects. Given an object \emph{o} and a set \emph{S}, one can only enquire whether $o \in S$. If one imposes order on the objects within a set then one obtains sequences; e.g. vectors, strings, permutations etc.. In a sequence $T$ if $x \in T$ then one can also query the position of $x$. A permutation on an alphabet $ \Sigma $, is a sequence where every object in $\Sigma $ occurs precisely once. In a string a symbol can repeat whereas in a permutation there is bijection from the positions to the symbols. Given a permutation $ \pi $= ($\pi_{1} $, $ \pi_{2} $, $ \pi_{3} $,.....,$ \pi_{n} $) over the alphabet $ \Sigma $ =$\{ $0, 1, . . . , n$-$1 $\}$  $ \pi_{i} $ and $ \pi_{i+1} $ form an \emph{adjacency} if $ \pi_{i+1} $ =$ \pi_{i} $+1, we call this as \emph{normal} adjacency or \emph{type 1} adjacency. The concept of adjacencies is widely used computation. The set of permutations with n symbols is called a symmetric group that we denote with P$_{n} $. The identity permutation with n symbols denoted by I$ _{n} $ where I$ _{n} $=(0,1,2,...,n$-$1) has exactly n$ - $1 adjacencies. Likewise, the reverse order permutation denoted by R$ _{n} $ where R$ _{n} $=(n$-$1, n$-$2, n$-$3, n$-$4, ...,0) has no adjacencies. We say that $\pi^a \in P_n(k)$ reduces to $\pi^b \in P_{n-k}(0)$ if  $\pi^b $ is obtained by eliminating all the adjacencies in  $\pi^a$. For example, (4, 5, 2, 1, 3, 0) in P$_6$ reduces to (4, 2, 1, 3, 0) in P$_5$  where  (4, 2, 1, 3, 0) is $irreducible$. The algorithm for reduction identifies and eliminates all maximal blocks of consecutive adjacencies ($>0$). Let the first symbol of one such block B be f and the last one be l then B is replaced by f and the value of every symbol with value $>$l is decreased by l$-$f. This process is repeated until all adjacencies are eliminated.\\
\hspace*{1cm} 
 In this article, $\forall k$ we compute the cardinalities of P$_{n} $(k) that is, we compute $ \forall_k \mid $P$_{n} $(k)$ \mid $ in O(n$ ^{2} $) time.  
We call the classic adjacency as type 1 adjacency. We define three variations of it.  The first variation is  back-adjacency or simply b-adjacency or type 2 adjacency where in addition to the normal adjacencies if $ \pi_{n} $ =n$ - $1 then it forms an adjacency with (an imagined) $ \pi_{n+1} $ =n. 
The second variation of adjacency is called front-adjacency or simply f-adjacency or type 3 adjacency where in addition to the normal adjacencies if $ \pi_{1} $ =0 then it forms an adjacency with (imagined) $ \pi_{0} $ =$ - $1. 
The third variation of adjacency is called front-and-back-adjacency or simply bf-adjacency or type 4 adjacency where in addition to the normal adjacencies if $ \pi_{n} $ =n$ - $1 then it forms an adjacency with (imagined) $ \pi_{n+1} $ =n and if $ \pi_{1} $ =0 then it forms an adjacency with (imagined)$ \pi_{0} $ = $ - $1.
P$ _{n} $(k) denotes the set of permutations in P$ _{n} $ with exactly k adjacencies; the type of adjacency will be evident from the context.
We compute $ \forall k~\mid $P$ _{n} $(k)$ \mid $  in O(n$ ^{2} $) time for any type of adjacency.   When necessary we employ the notation P$ _{n} (k, i)$ where $i$ indicates the type of adjacency.  We call two permutations $\pi^a$ and $\pi^b$ as $mirrors$ of each other if they are corresponding permutations from two different alphabets. That is, (0, 2, 1) and (1, 3, 2) are mirrors of each other. Mirrors are equivalent; i.e. the numbers of adjacencies and the numbers of moves that are required to sort them are identical.

\hspace*{1cm}The concept of adjacencies is inherent in sorting by comparison algorithms. Quicksort seeks to reduce the inversions in a permutation by swapping two distant objects that form an inversion whereas bubble sort swaps adjacent objects that form an inversion and thus, it reduces exactly one inversion per swap. Adjacencies and inversions are inherently related. An inversion exists if and only if the total number of (type 1) adjacencies is less than n$ - $1. 
All the algorithms terminate when n b-adjacencies are created \cite {cormen}. 
Graham et al. study the related topics: ascents, cycles, left-to-right maxima and excedances in permutations \cite {graham}. Given $ \pi $= ($\pi_{1} $, $ \pi_{2} $, $ \pi_{3} $,....., $ \pi_{n} $) an ascent is defined as a position j where $ \pi_{j} $$ < $$ \pi_{j+1} $ and the corresponding recurrence relation is given by $ \alpha $(n,k)= $ \alpha $(n$-$1,k)*(k+1) + $ \alpha $(n$-$1,k$-$1)*(n$-$k) \cite {euler, graham, shan}.  $ \alpha(n,k)$ denotes the number of permutations in $P_n$ that have exactly k ascents.
The numbers thus generated are known as Eulerian numbers. The cycles in permutations correspond to Stirling numbers of the first kind \cite {stirling, graham}. The left to right maxima corresponds to $ \pi_{j} $ where $ \forall_{i<j} $ $ \pi_{i} $$ < $$ \pi_{j} $ \cite {graham}. A symbol $ \pi_{j} $ in $ \pi $ is an excedance if j$< \pi_{j} $ \cite{graham}.

\hspace*{1cm}Transforming permutations with transpositions, and prefix transpositions has been well studied. The symmetric $distance$ between two permutations $\alpha$ and $\beta$ with a symmetric operation $\tau$ i.e. $d_{\tau} (\alpha, \beta)$ is the minimum number of $\tau$ operations required to transform $\alpha$ into $\beta$ or vice versa. So, the transposition distance between $\alpha$ and $\beta$, i.e.  $d_{T} (\alpha, \beta)$ is the minimum number of transpositions required to transform  $\alpha$ into $\beta$ or vice versa.
The concept of \emph{breakpoint} was used in many articles, e.g. \cite {bafna, dias}  where a breakpoint denotes an absence of an adjacency. 
Bafna and Pevzner \cite {bafna} studied sorting permutations in P$ _{n} $ with transpositions and showed a lower bound of $ \lfloor n/2 \rfloor $+ 1 and an upper bound of $ \frac{3n}{4} $. They also gave a 1.5 approximation algorithm for the same.   
Eriksson et al. improved the upper bound to $ \frac{2n}{3} $ \cite {eriksson} and also showed that R$ _{n} $, the reverse order permutation can be sorted in $ \frac{n+1}{2} $ transpositions.

\hspace*{1cm} Dias and Meidanis \cite{dias}studied the prefix transposition distance over P$_n$ and showed that: (a) n$-$1 is an upper bound, (b)$ \frac{n}{2} $ is  a lower bound, and (c)  R$ _{n} $, can be sorted in $\frac{3n}{4} $ prefix transpositions. They conjectured that R$ _{n} $ is the hardest permutation to sort. Recall the R$ _{n} $ has no adjacencies. Chitturi and Sudborough improved the lower bound to $\frac{3n}{4} $ \cite {chitturi08} and the upper bound to n-$\log_{\frac{9}{2}}n$ \cite {chitturi}. Labarre \cite {labarre} improved the lower bound of prefix transposition distance over P$ _{n}$ to $\frac{3n}{4}$. Recently, Chitturi \cite{chitturi15} showed that an upper bound for the prefix transposition distance over P$_n$ is $n-\log_{\frac{7}{2}}n$. 

\hspace*{1cm} Sorting permutations with prefix reversals i.e. flips, also known as the pancake problem has been widely studied. The best known upper bound for this problem is $18n/11 + O(1)$ \cite{ChitturiFMMSSV09}. Cibulka showed that sorting a random stack of n pancakes can be done with at most $\frac{17n}{12} $+O(1) flips on average. The average number of flips of the optimal algorithm for sorting stacks of n burnt pancakes is shown to be between n+$ \Omega $$ (\frac{n}{logn}) $ and $\frac{7n}{4} $+O(1) and the author conjectures that it is n+$ \Theta $$ (\frac{n}{logn}) $ \cite{cibulka}.

\hspace*{1cm}Bulteau et al. show that sorting permutations by transpositions is NP-hard \cite {bulteau}. Thus, it is desirable to estimate the expected number of moves to sort a permutation $ \in $P$ _{n} $ with transpositions. It is believed that sorting permutations by either prefix or suffix transpositions  is intractable. So, the model that estimates the expected number of moves to sort permutations with various block-moves is sought.

\hspace*{1cm}The block moves are studied on strings as well. The earliest known articles on transforming strings with transpositions and prefix transpositions are \cite{christieI01} and  \cite{chitturi07, chitturi, chitturi08} respectively. Adjacent transpositions, where adjacent elements swap their positions, on permutations has a proven exact upper bound \cite{AkersK89, chitturi13, LakshmivarahanJD93} whereas an efficient algorithm to count the exact number of adjacent transpositions required to transform one string to another string is presented in \cite{chitturiSVF08}. Further, cyclic \emph{short swaps} and adjacent transpositions are studied on permutations in \cite{feng10}.

\hspace*{1cm}The main contributions of this article are: (i)  computing $ \forall_{k} $  $ \mid $P$ _{n} $ (k)$ \mid $ for any type of adjacency in O($n^2$) time, 
(ii) a theoretical framework that forms a basis for models that estimate the expected number of block-moves to sort a permutation in P$_n$(0) (and thus, in P$ _{n} $).
We were made aware of OEIS and \cite{whitworth} by an anonymous referee which lead to \cite{Tanny}. An examination of some intger sequences in OEIS reveals that our article provides an alternative explanation for some of the known integer sequences.
To our knowledge, the current type of exploration of adjacencies in permutations and their applications are novel.

\section{Regular adjacencies}
\onehalfspacing
Symbols $ \pi_{i} $ and $ \pi_{i+1} $ form an adjacency if $ \pi_{i+1} $ =$ \pi_{i} $+1 in regular i.e. type 1 adjacencies. The sorted permutation with n symbols, i.e. $\textbf{I$_{n}$}$ where $ \forall_{k} $$ \pi_{k} $ = k, also called as the identity permutation, has the maximum number i.e. n$ - $1 adjacencies. The reverse order permutation i.e. R$ _{n} $ where $ \forall_{k} $  $ \pi_{k} $ = n$ - $k, has zero adjacencies. A permutation is \emph{reduced} or \emph{irreducible} if it has no adjacencies \cite {chitturi}. Let $ \pi $ a member of P$ _{7} $ be (4 6 3 1 2 0 5) then (3 5 2 1 0 4) is the reduced form of $ \pi $ where (1 2) is reduced to 1 and all the symbols with a value greater than two are decremented by one. This process is repeated until all the adjacencies are removed from $\pi$. Here, the resulting permutation is a member of P$ _{6}$. 

The following theorem establishes a recurrence relation to compute $ \mid $P$ _{n} $(k)$ \mid $ for the first type of adjacencies. 
\newtheorem{theorem}{Theorem}
\newtheorem{obs}{Observation}
\newtheorem{corollary}{Corollary}[theorem]
\begin{theorem}
\label{th_type1_adj}
\normalfont
 Let P$ _{n} $(k) be the subset of P$ _{n} $ where any $ \pi $$ \in $ P$ _{n} $(k) has exactly k type 1 adjacencies. Let f(n,k) be the cardinality of P$ _{n} $(k). Then f(n,k) = f(n$ - $1,k$ - $1)+ (n$ - $1$ - $k) * f(n$ - $1,k) + (k+1)* f(n$ - $1,k+1) where $ 0 \leq $k$<$n.
\end{theorem}
%
%
%
 The cardinalities of $P_0(n)$, $P_1(n)$ etc. occur in OEIS  \cite{oeis} with sequence numbers A000255, A000166 etc.. 
Tanny studied the cardinalities of the sets of permutations with n symbols and k successions (or type 1 adjacencies ) \cite{Tanny}.
He gave the expression for f(n,k) as follows where  D$ _{i} $ is a derangement number for size $i$ \cite{Tanny}.
f(n,k)=$\begin{pmatrix}
     n-1  \\ 
   k \\
\end{pmatrix}$ (D$ _{n-k} $ +D$ _{n-1-k} $). 

Tanny also studied circular successions where $\pi_i$ and $\pi_{\equiv (i+1)}$ form an adjacency if $\pi_{\equiv (i+1)} \equiv (1+ \pi_i) $. Here $\equiv x$ denotes $ x \mod{n} $. 
He showed that 
$\lim_{n \rightarrow \infinity}$ (Q$ ^{*}(n,k)/n! $)=$ e^{-1} /k !$ where Q$ ^{*}(n,k)$ denotes number of permutations with k circular successions \cite{Tanny}.

Roselle determined the cardinality of P(n,r,s) as P(n,r,s)=$\begin{pmatrix}
     n-1  \\ 
   s \\
\end{pmatrix}$ P(n-s,r-s,0) where  s denotes the number of type 1 adjacencies (that he calls successions) and r denotes the number of rises \cite{Roselle}. A rise in a permutation exists at a position $i$ if $ \pi_{i} $$ < $$ \pi_{i+1} $ \cite{Roselle}.\\
 
\section{ Adjacency Variations}
\subsection{ Type  2  adjacency}
In type  2 adjacency or b-adjacency, in addition to the adjacencies in type 1 adjacency we imagine that $ \pi _{n+1}$= n; i.e. if $ \pi _{n}$= n$ - $1 then $ \pi _{n}$ and $ \pi _{n+1}$ form an adjacency.
The sorted permutation with n symbols, i.e. \textbf{I$ _{n} $}  where $ \forall_{k} $ $ \pi _{k}$ =k, also called as the identity permutation, has the maximum number i.e. n adjacencies.
The reverse order permutation i.e. \textbf{R$_{n}$}  where $ \forall_{k} $ $ \pi_{k} $ =n$ - $k, has zero adjacencies.
If $ \pi $ = (4 6 3 5 0 2 1 7) then (4 6 3 5 0 2 1) is the reduced form of it where $ \pi_{n} $ is deleted because $ \pi_{n} $= n$-$1. 

\begin{theorem}
\normalfont
\label{th_type2_adj}

 Let f(i, j) denote the number of permutations in P$ _{i} $ with exactly j adjacencies. Then the
recurrence relation for f(i, j) is:\\
f(i, j)= (f(i$-$1, j$-$1)$-$f(i$-$2, j$-$2))*2 +f(i$-$2, j$-$2) +\\
(f(i$-$1, j+1)$-$f(i$-$2, j))*(j+1)+f(i$-$2, j)*(i$-$j$-$1)+\\
(f(i$-$1, j)$-$f(i$-$2, j$-$1))*(i$-$j$-$2) +\\
 f(i$-$2, j+1)*(j+1); 0$ \leq $ j$ \leq $i+1.			
 \end{theorem}

\subsection{ Type 3 adjacency}
In type  3 adjacency or f-adjacency, in addition to the adjacencies in type  1 adjacency we imagine that $ \pi_{0} $=$ - $1. That is, if $ \pi_{1} $= 0 then $ \pi_{0} $ and $ \pi_{1} $ form an adjacency. \textbf{I$ _{n} $} has the maximum number i.e. n adjacencies and  \textbf{R$_{n}$}  has zero adjacencies.
Type  2 and type  3 adjacencies are symmetrical.In addition to the adjacencies defined in type 1 adjacency, an adjacency is defined between $ \pi_{n} $ and $ \pi_{n+1} $ in b-adjacency whereas the same is defined between $ \pi_{0} $ and $ \pi_{1} $ in the f-adjacency.
The recurrences governing $\forall k \mid $P$ _{k} $(n)$ \mid $ and their base values (i.e. $ n \leq 4$) are identical for type 2 and type 3 adjacencies. Thus, yielding identical values for  $ \mid $P$ _{k} $(n)$ \mid $ for f-adjacency and b-adjacency.

 The cardinalities of $P_0(n)$, $P_1(n)$ etc. occur in OEIS \cite{oeis} with sequence numbers A000166 denoting subfactorial or rencontres numbers, or derangements: number of permutations of n elements with no fixed points; A000240: rencontres numbers: number of permutations of [n] with exactly one fixed point etc..

\subsection{ Type 4 adjacency}
Type 4 adjacency or bf-adjacency has additional adjacencies defined compared to  type 1 adjacency. We imagine that $\pi_{n+1}$= n and $\pi_0=-1$.
That is if $ \pi_{n} $= n$ - $1 then $ \pi_{n} $ and $ \pi_{n+1} $ form an adjacency; likewise, if $ \pi_{1} $=0 then $ \pi_{0} $ and $ \pi_{1} $ form an adjacency.
 \textbf{I$ _{n} $} has the maximum number of adjacencies, i.e. n+1 and R$ _{n} $ has zero adjacencies. If $ \pi $ = (0 4 6 3 5 2 1 7) then (4 6 3 5 2 1) is the reduced form of it where $ \pi_{n} $ is deleted because $ \pi_{n} $= n$ - $1 and $ \pi_{1} $ is deleted because $ \pi_{1} $= 0. The following theorem establishes a recurrence relation to compute $ \mid $P$ _{n} $(k)$ \mid $ for bf-adjacency.
 
 The cardinalities of $P_0(n)$, $P_1(n)$ and $P_2(n)$  occur in OEIS \cite{oeis} with the following sequence numbers.
  A000757: Number of cyclic permutations of n symbols with no [i] immediately followed by [i+1] where [i] denotes $i \% n$; A135799: second column (k=1) of triangle A134832 (circular succession numbers);  A134515: third column (k=1) of triangle A134832, etc..
%
\begin{theorem}
\normalfont
\label{th_type4_adj}

 Let f(i, j) denote the number of permutations in P$ _{i} $ with exactly j adjacencies. Then the
recurrence relation for f(i, j) is:\\
f(i, j)= (f(i$-$1, j)$-$f(i$-$2, j$-$1))*(i$-$j$-$2) +\\
(f(i$-$1, j$-$1)$-$f(i$-$2, j$-$2))*2 +f(i$-$2, j$-$2) +\\
(f(i$-$1, j+1)$-$f(i$-$2, j))*(j+1)+f(i$-$2, j)*(i$-$j$-$1)+\\
 f(i$-$2, j+1)*(j+1); 0$ \leq $ j$ \leq $i+1.			
 \end{theorem}

\section{A General Model for Block Moves}
Transpositions, prefix transpositions and suffix transpositions are called as  block-moves in this article. In the sequel a move refers to any of the above operations and its meaning is clarified by the context. If we are referring to a particular operation then the context clarifies the same. We assume that the permutation that we are sorting is irreducible. A prefix transposition can either create or destroy zero, one or two adjacencies. Likewise, a transposition can either create or destroy zero, one, two or three adjacencies. The moves that create one, two or three adjacencies are called a \emph{single}, a  \emph{double} and a  \emph{triple} respectively. A move can also break adjacencies, however, because the permutation that is being sorted is irreducible we do not consider such moves.\\ 
Bulteau et al. show that sorting permutations by transpositions is NP-hard \cite {bulteau}. Thus, it is desirable to
estimate the expected number of moves to sort a permutation $ \in $P$ _{n} $ with transpositions. It is believed that sorting permutations by either prefix transpositions or suffix transpositions is also hard. So, a model that estimates the expected number of moves to sort permutations with various block-moves is sought.
\\\hspace*{1.5cm}The first type of adjacencies can be used for estimating the expected number of moves to sort a permutation with any  block-move, e.g. transposition, prefix transposition, suffix transposition, prefix/suffix transposition. However, the other types of adjacencies are more apt depending on the operation. For example, f-adjacency is applicable to prefix transpositions and the b-adjacency is applicable to suffix transpositions. The bf-adjacency is applicable for transpositions.
\\\hspace*{1.5cm}Christie \cite {christie} showed that $d_t( \pi ,I _{n} )= d_t( \pi^{*} ,I_{n} )$ where $d_t(A, B)$ is the transposition distance between A and B and $ \pi^{*} $ is the reduced form of $ \pi $. Similar to an optimal sequence of transpositions that sort a given permutation, an optimal sequence of prefix transpositions also need not break any adjacencies. It follows
that $d_{pt}(\pi ,I_{n})= d_{pt}(\pi^{*},I _{n})$.
\vspace*{0.2cm}
\begin{theorem}
\normalfont
\label{th_cardinality_p_n_k}
 Let S be the maximal set of permutations in P$_{n}(n-k)$ such that every permutation in S yields the same permutation in $P_k(0)$ upon reduction. If $ \mid S \mid = \mu$ then for any given permutation $\pi^x \in$ P$_{k}(0)$  there are exactly $\mu$ permutations in $P_n$ that reduce to $\pi^x$.
\end{theorem}
$ \textbf{Proof:} $ Consider P$ _{n} $(n$ - $k), the set of all permutations $ \in P_n$ whose reduced length is k. That is, for each $ \pi \in$ P$ _{n} $(n$ - $k) there is a corresponding permutation  $ \pi^* \in$ P$ _{k} $(0).  The numbers of the corresponding transposition based moves to sort $\pi$ and $\pi^*$ are identical. Thus, we treat them as equivalent.
 Under this equivalence, we seek to show that P$ _{n} $(n$ - $k) is a multiset composed exclusively of some $c \in Z^{+}$ copies of each $\pi \in$ P$_{k}(0)$.
  
 We first analyze  type 1 adjacency. Consider P$_{n}$(n$-$k) where n=3 and k=3 it consists of three permutations: $\{(0 2 1), (1 0 2),(2 1 0) \}$ corresponding to P$ _{3}$(0). 
 Consider a split of I$ _{5} $ =(0 1 2 3 4) into three substrings s$ _{1} $, s$ _{2} $, s$ _{3} $ such that $ \forall_{i} $ $ \mid $s$ _{i} $$ \mid $$ > $0 yielding I$ _{5}^{*} $ where I$ _{5} ^{*}$= (0 1, 2 3, 4). Here we separate adjacent substrings with comma. Note that in its reduced form I$ _{5}^{*} $ equals I$ _{3} $.
Consider an alternate split of I$ _{5} $ =(0 1 2 3 4) into three substrings t$ _{1} $, t$ _{2} $, t$ _{3} $ yielding I$ _{5}^{*} $=(0, 1, 2 3 4).
Let A=$\{$(0 1 4 2 3), (2 3 0 1 4),(4 2 3 0 1) $\}$ and B=$\{$(0 2 3 4 1), (1 0 2 3 4),(2 3 4 1 0) $\}$. 
A and B belong to I$ _{5} $ where  where both these sets in their reduced form equal $\{$(0 2 1), (1 0 2),(2 1 0)  $\}$ which is the same as P$ _{3} $(0). 
That is, for each distinct split of (0 1 2 3 4) one will have three permutations in P$ _{5} $. The number of such splits equals the number of integer solutions to the equation x1 + x2 +x3=5 i.e. $ {5-1} \choose {3-1} $. 
Extending this argument to a general k and a general n (n$ > $k), there are $ {n-1} \choose {k-1} $ copies of P$ _{n} $(n$-$k) in P$ _{n} $. That is, any member of P$ _{k} $(0) has exactly $c=$ $ {n-1} \choose {k-1} $ occurrences in P$ _{n} $.

\hspace*{1.5cm} Consider  type 2 adjacency. Here, P$ _{3} $(0) =$\{$(0 2 1), (2 1 0) $\} $. Consider a split of I$ _{5} $ =(0 1 2 3 4) into three substrings s$ _{1} $, s$ _{2} $, s$ _{3} $ such that $ \forall_{i} $ $ \mid $s$_{i} $$ \mid $$ > $0 to yield I$ _{5}^{*} $ where I$ _{5}^{*} $= (0 1, 2 3, 4). Note that I$ _{5}^{*} $ is equivalent to I$ _{3} $. Consider an alternate split of I$ _{5} $ =(0 1 2 3 4) into t$ _{1} $, t$ _{2} $, t$ _{3} $ yielding I$ _{5}^{*} $=(0, 1, 2 3 4). 
Let   A=$\{$(0 1 4 2 3), (4 2 3 0 1) $\} $ and B=$\{$(0 2 3 4 1), (2 3 4 1 0) $\} $.
A and B belong to I$ _{5} $ where where both these sets in their reduced form equal C, where C= $\{$(0 2 1), (2 1 0) $\} $ which is the same as P$ _{3} $(0). 
First we note that if $n-1$ is in the last position then if a permutation from P$_{n-1}(n-k-1)$ precedes it then effectively we have a permutation in  P$_{n}(n-k)$. In the above example consider a split of (0 1 2 3) into three non empty substrings, say (0 1, 2,  3). This will yield (3 2 0 1 4) as a permutation in I$_5$(2) where the trailing 4 remains in the last position.
Thus, extending the above argument to a general k and a general n (n$ > $k), there are ${n-1} \choose {k-1}$ $+b$ copies of P$ _{n} $(n$-$k) in P$ _{n} $ where $b$ is the number of copies of  P$ _{n-1} $(n$-$k-1) in P$ _{n-1} $. If one expands the recurrence then one obtains the total number of copies of of P$ _{n} $(n$-$k) in P$ _{n} $ as $ c= \sum_{n=1}^{n-k+1} {n-i \choose {k-1} }$
That is, any member of P$ _{k} $(0) has exactly $ c $ occurrences in P$ _{n} $.
Likewise, the same can be shown from f-adjacency. \\
\hspace*{1.5cm}Consider  type 4 adjacency. Here, P$ _{3} $(0) =$\{$ (2 1 0) $\} $. Consider splitting of I$ _{5}$ =(0 1 2 3 4) into three substrings s$ _{1} $, s$ _{2} $, s$ _{3} $ such that $ \forall_{i} $ $ \mid $s$ _{i} $$ \mid $$ > $0 yielding I$ _{5}^{*}$ where I$ _{5}^{*}$= (0 1, 2 3, 4). Note that I$ _{5}^{*} $ is equivalent to I$ _{3} $.
  Consider an alternative split of I$ _{5} $ =(0 1 2 3 4) into t$ _{1} $, t$ _{2} $, t$ _{3} $ yielding I$ _{5}^{*} $=(0,
1, 2 3 4). Note that the sets of permutations A and B belong to I$ _{5} $ where A=$\{$(4 2 3 0 1)$\} $ and B=$\{$ (2
3 4 1 0) $\} $ where both these sets in their reduced form are $\{$(2 1 0)$\} $ which is the same as P$ _{3} $(0). 

Let $\pi \in P_n$ and let $\pi_{n} =n-1$. We note that if a permutation from P$_{n-1}(n-k-1)$ precedes $\pi_{n}$ then effectively we have a permutation in P$_{n}(n-k)$. Likewise, if $\pi_{1} =0$ is in the first position then if a permutation from P$_{n-1}(n-k-1)$ succeeds it then effectively we have a permutation in  P$_{n}(n-k)$. Note that here a permutation from P$_{n-1}(n-k-1)$ is a mirror permutation that is it is defined on the alphabet $(1,2,\ldots n-1)$.
However, the above cases count the permutations that begin with 0 and end with $n-1$ twice. The number of such permutations is $\mid$ P$_{n-2}$(n-k-2) $\mid$.  
So, the recurrence is
$\mid$P$_{n}(k)\mid$ = $ 2* \mid$P$_{n-1}(n-k-1) \mid -$  $\mid $ P$_{n-2}$(n-k-2) $\mid$.
This recurrence relation has coefficients that are positive integers and the base case for this recurrence relation is $\mid $ P$_{k}$(0) $\mid$ where the base case corresponds to one copy of P$_{k}$(0). Thus, we can conclude that there are 
integral number of copies of P$ _{k} $(0) in P$ _{n} $(n$-$k).$~\blacksquare$\\


We define the set of irreducible permutations in P$_n$ as the \emph{vector alphabet} of P$_n$ and it is denoted by $\alpha$(P$_n$). Note that $\alpha$(P$_n$) is a set i.e. if any permutation $\pi^x \in P_n$ reduces to an irreducible permutation $\pi$ then only $\pi$ will be a member of $\alpha$(P$_n$).
Let the \emph{offset} denoted by $\delta$ be the term that must be added to n to obtain the  maximum number of adjacencies possible for each type of adjacency. That is, $\delta = -1$ for type 1 adjacency, $\delta = 0$ for type 2 and type 3 adjacencies, and $\delta = 1$ for type 4 adjacency. Furthermore, let  {P$_k$(0)}$^{c_k}$ denote $c_k$ copies of the set P$_k$(0) where $c_k \in Z^+$. The corollaries given below follow.
%
%
\begin{corollary}
\label{corr_cardinality_alphabet}
\normalfont
$\alpha$(P$_n$)= $\bigcup\limits_{k=1}^{n+\delta}$ P$_k$(0). $~\blacksquare$\\
\end{corollary}
%
%
\begin{corollary}
\label{corr_cardinality_composition}
\normalfont
 P$_n$= $\bigcup\limits_{k=1}^{n+\delta}$ {P$_k$(0)}$^{c_k}$. $~\blacksquare$\\
\end{corollary}
\begin{corollary}
\label{corr_cardinality_exp_moves}
\normalfont
 Let $ \phi $(P$ _{n} $(k)) denote the average number of moves to optimally sort all permutations
in P$ _{n} $(k) with block moves. If $ \phi $(P$ _{k} $(0))=$ \mu $ then $ \phi $(P$ _{n} $(n$ - $k))= $ \mu $ for n$ > $k.$~\blacksquare$
\end{corollary}
%
%
%
\section*{Model}
 Theorem \ref{th_type1_adj} and Theorem \ref{th_type4_adj} lead to the corresponding algorithms that compute $\forall{_k} $ P$ _{n}(k) $. Theorem \ref{th_cardinality_p_n_k} and Corollary \ref{corr_cardinality_exp_moves} show that the distribution of P$_k$(0) is uniform in P$ _{n} $ for all n$ > $k. That is, every $\pi \in$  P$ _{k}(0)$ has exactly some c ($\in Z^+$) permutations in P$ _{n}$ that reduce to it.
 For each of the operation we evaluate the probabilities of executing a single, a double, and a triple i.e. $p_1,~p_2,~p_3$ on a (uniformly) random permutation in $P_{n}(0)$. A prefix or a suffix transposition does not admit a triple. Based on these probabilities, we compute the expected number of adjacencies created per one move in P$_n$, $\psi$. 
 We employ $\psi$, the limiting value of $\psi$  and the expected/estimated number of moves to optimally sort a permutations in P $_{2} $(0) $ \ldots $  P$ _{i-1} $(0) to compute the estimated number of moves that are required to sort a $\pi \in$ P$ _i $(0). 
 For example it can be seen that the limiting value of $\psi$ for prefix transpositions is 1.5 from Observation \ref{obs_ptexpadj}.  The goal of sorting a permutation is to obtain a permutation of size one (after reduction) by starting with a permutation in P$_n(0)$. Thus, our first measure for expected number of moves to sort a permutation in P$_n(0)$ is $(n-1)/\psi$.
The second measure computes the weighted average of the estimates for $P_{n-x}(0)$ and $P_{n-x+1}(0)$ and adds one to it where $n-x< n-\psi < n-x+1$.  The weighted average is based on the position of $n-\psi$ in $[n-x, n-x+1]$.
In this measure, in one move the size of the permutation is presumed to be reduced by $\psi$ and we add the expected number of moves to sort the permutation of the resultant size. Note that this is not an integer size, so, we compute the weighted average.
The first measure mimics the future behavior of the expected number of moves and the second measure mimics the past behavior of the expected number of moves.  We take the mean of the above two measures as the estimate for the expected number of moves to sort a permutation in P$_n(0)$.
 
The following algorithm $\textit{Move\_Count} $ estimates the expected number of moves to sort a permutation $ \pi $ $ \in $P$ _{n} $(0). 
The average number of moves to sort a permutation $ \pi $ $ \in $P$ _{i} $(0) for i=(2..limit) is computed by a branch and bound program. These values are used as base cases for \textit{Move\_Count}.
\newline
Algorithm Move\_Count(i)
\\
{\small
Precomputation. Execute a branch and bound algorithm that computes the average number of moves to sort for all
permutations $ \pi $$ \in $P$ _{n}(0)$ for n=2,....,limit. Let base[2..limit] hold the respective averages. 
}
{\small
\begin{algorithmic}

\State Intilization: cnt=0
\For{(i=limit+1,....max)}
\State j $ \gets$ i
\State j$ \gets $ j$-\psi(n) $  \hspace{1cm}    $ \setminus\setminus $ Observation \ref{obs_ptexpadj},   $\psi(n)$: expected number adjacencies that a move in $P_n$ creates
\State x $ \gets $ 1 + (j$ - \lfloor j \rfloor$)* base[$\lceil j  \rceil$] +($\lceil j \rceil-$j)*base[$\lfloor j \rfloor$]

\State y $ \gets $ $(n-1)/\psi$
\State base[i] $ \gets $ $(x+y)/2$

 \EndFor
\end{algorithmic}
}
We let X be the random variable that denotes the expected number of moves to sort a $ \pi $$ \in $ P$ _{n} $. Due to Theorem \ref{th_cardinality_p_n_k}, E(X)= $ \forall_{i} $$ \Sigma $ f$ _{i} $ E(X$ _{i} $) where X$ _{i} $ is the random variable that denotes the expected number of moves required to sort a $ \pi $$ \in $ P$_n (i) $  and f$ _{i} $=$ \frac{|P_{n}(i)|}{n!} $. 
The estimate for E(X) is evaluated by algorithm Expected\_Value. Note that we use the appropriate definition for adjacency, and the corresponding algorithm Adjacency\_Countx is used (for the type of adjacency x, where x $\in$1, 2, 4).\\
\\Algorithm Expected\_value

{\small
\begin{algorithmic}

\State Intilization: for ( j= 2 ...limit) set E(X$ _{j} $) from the branch and bound program.
\State offset that determines the maximum possible adjacencies =$ - $1 for x=1, =0 for x=2 or 3, =+1 for x=4. 
\For{(i=limit+1,....n)}
compute E(X$ _{j} $)by executing Move\_Count(j). $ \setminus\setminus $Note that this order is important
 \EndFor
 \For{(i=limit+1,....n)}
 \For{(i=0...j+offset)}
  f$_{i} $=$\frac{|P_{j}(i)|}{j!} $  $ \setminus\setminus $ $ |P_{j}(i)| $ is read from output of the appropriate Adjacency\_Countx
   \EndFor \\
  Estimate of E(X) = $ \Sigma $ f$ _{i} $E(X$ _{i} $).
  \EndFor
\end{algorithmic}
}
Algorithm Move\_Count uses the already computed averages for the number of moves required to sort all permutations with zero adjacencies of a given size, up to size eight. These numbers are used as base cases to compute the same for larger values of n.
\\The following theorem establishes a lower bound for the fraction of permutations in P$ _{n} $ that have exactly one adjacency. Note that a permutation $ \pi $$ \in $P$ _{n} $ (k) where k$ \geq $1 can be reduced to $ \pi ^{*}$$ \in $P$ _{n-k} $.
Thus, if the results for optimally sorting all permutations for P$ _{i} $ (i$ < $n) are known then one need only
look up the results for $ \pi^{*} $. So, the computation of optimal sequences is required only for irreducible
permutations. From the following theorem it follows that for P$ _{n} $, approximately $ \frac{n!}{e} $ permutations require the computation of optimal moves.\\
\begin{obs}
\label{th_inequality}
\normalfont
Let f$ _{i} $ (0)= $ \frac{|P_{n}(0)|}{n!} $ where i is the type of adjacency $ \in $ $ \{ $1,2,3,4 $\}$ and $ \mid $P$_{n}$(0)$ \mid $ be the corresponding magnitude of the set of irreducible permutations $ \in $P$ _{n} $. We have the following inequalities:(i) f$ _{1} $(0)$ > $$ \frac{1}{e} $. (ii) f$ _{2} $(0)$ \leq $$ \frac{1}{e} $. (iii) f$ _{3} $(0)$ \leq $$ \frac{1}{e} $.(iv) f$ _{4} $(0)$ < $$ \frac{1}{e} $.
\end{obs}
$ \textbf{Proof:} $ Consider the first type of adjacency. Recall that $ \Sigma $=$\{$0,1,2,..., n$ - $1$\}$. The probability that n$ - $1 is not present at a given position is $ \frac{n-1}{n} $. Given that that n$ - $1 does not occur at position i, the probability that an adjacency exists between $ \pi_{i} $ and $ \pi_{i+1} $ is $ \frac{1}{n-1} $, note that out of n$- $1 remaining
symbols only $ \pi_{i+1} $ is favorable for position i+1. Thus, the probability that there is an adjacency between $ \pi_{i} $ and $ \pi_{i+1} $ is = $ \frac{n-1}{n} $ * $ \frac{1}{n-1} $ =$ \frac{1}{n} $. Thus, the probability that there is no adjacency between $ \pi_{i}$  and $ \pi_{i+1} $ is 1$ - $ $\frac{1}{n} $. So, the probability that there is no adjacency between $ \pi_{i} $ and $ \pi_{i+1} $ for 1$ \leq $i$ \leq $n$ - $1, that we call p$_{1} $(0)=(1$ - $$\frac{1}{n} $) $ ^{n-1} $ =$ \frac{(1-\frac{1}{n})^{n}}{(1-\frac{1}{n})} $ . 
For large values of n, p$ _{1} $(0) $ \approxeq \frac{1}{e*(1-\frac{1}{n})} $. Thus, $ \frac{1}{e} $ is a strict lower bound.\\
\hspace*{1.5cm}
Consider the second type of adjacency. Similar to the first type of adjacency, the probability
that there is no adjacency between $ \pi_{i} $ and $ \pi_{i+1} $ for 1$ \leq $i$ \leq $n$ - $1=(1$ - $$ \frac{1}{n} $)$ ^{n-1}$. Additionally , the probability
that $ \pi_{n} $ $ \neq $n$ - $1 is (1$ - $$ \frac{1}{n} $). So, p$ _{2} $(0), the probability that no adjacency exists is (1$ - $$ \frac{1}{n} $)$ ^{n} $. For large values of n, p$ _{2} $(0) $ \approxeq \frac{1}{e} $. Likewise, p$ _{3} $(0) $\approxeq \frac{1}{e} $. Note that instead of $ \pi_{n} $ $ \neq $n$ - $1 here we require that $ \pi_{1} $ $ \neq $0 with the identical probability. Thus, p$_{2}$(0)$ \leq $ $ \frac{1}{e} $ and p$ _{3} $(0)$ \leq $$ \frac{1}{e} $.
 
Consider the fourth type of adjacency. In addition to the restrictions of second type of
adjacency we require that $ \pi_{1} $ $ \neq $0. This yields p$ _{4} $(0)=(1$ - $$ \frac{1}{n} $)$^{(n+1)}$$ \approxeq \frac{(1-\frac{1}{n})}{e} $ . Thus, p$ _{4} $(0)$ < $ $ \frac{1}{e}.$ $~\blacksquare$
\\

A similar result was given by Whitworth in \cite{whitworth}.
Whitworth evaluated the number of permutations without any adjacencies( of type 2 or type 3 ) as $ n!(e _{n}^{-1})$ \cite{whitworth}. Where $e _{n}^{-1}$ denotes the summation of the first n+1 terms in the series expansion of $e^{-1}$.

The following observation directly follows from Observation \ref{th_inequality}. It states that the computation of the expected number of moves for $P_n(0)$ is the bottleneck in the computation of the same for $P_n$, where $|P_n(0)|$ is equal to or approximately equal to $|P_n|/e$ for large values of n.

\begin{obs}
\label{th_inequality_corr}
\normalfont
Let the number of moves of a particular block-move to optimally sort any permutation in P$_i ~\forall~ i \leq n-1$ be known. For computing the expected number of moves to sort P$_{n} $ with the same operation, one needs to compute the optimum moves
(instead of looking up the answer) for approximately $ \frac{n!}{e}$ permutations. $~\blacksquare  $
\end{obs}
Note that even for bf-adjacency $ \forall_{n} $ $ \geq $ 20 the lower bound on $ \frac{P_{n}}{n!} $$ \geq $ 0.34056 and when $ \forall_{n} $ $ \geq $ 50 we
have  $ \frac{P_{n}}{n!} $$ \geq $ 0.35688. For larger values of n this fraction approaches (but never equals), $ \frac{1}{e}= 0.36787\ldots$.

\section{Prefix Transpositions }
A transposition exchanges adjacent sublists; when one of the sublists is restricted to be a prefix then one obtains a prefix transposition. Given a permutation $\pi$ that must be sorted, if $\pi_{n}=n-1$ then in an optimal sorting sequence $\pi_{n}$ need not be moved again. Likewise, if $\pi_{i}=i-1 ~\forall_{(i=n \ldots k)}$ then the last n-k+1 elements need not be moved again. This follows from a result in sorting transpositions by Christie \cite{christie}. Thus, type 2 adjacencies capture the adjacencies created by prefix transpositions. Likewise, type 3 adjacencies capture the adjacencies created by suffix transpositions and type 4 adjacencies capture the adjacencies created by transpositions.

A prefix transposition can either create or destroy zero, one or two adjacencies. The moves that create one, two or three adjacencies are called a single, a double and a triple respectively. A move can also break adjacencies, however, because the permutation that is being sorted is reduced no adjacencies exist, so, we do not consider such moves. The following two observations are well known.\\
\begin{obs}
\normalfont
\label{obs_ptsingle}
If $ \pi$ $\in$ P$_{n}$ (n$ > $1) is reduced then a single can always be executed.
\end{obs}
$ \textbf{Proof: } $ Let  $ \pi $= $\pi_{1} $, $ \pi_{2} $, $ \pi_{3} $,....., $ \pi_{n} $.\\
If $ \pi_{1}=0 $ then moving $ \pi_{1}$ to just before 1 creates a new adjacency. If $ \pi_{1}=n-1 $ then moving $ \pi_{1}$ to just after n-2
creates a new adjacency. If $ \pi_{1}\neq0 $  and $ \pi_{1}\neq n-1 $   then moving $ \pi_{1}$  to just before $ \pi_{1}+1$  or just after $ \pi_{1}-1$  creates a new adjacency. Note that these moves are both transpositions and prefix transpositions.$~\blacksquare$ \\
\begin{obs}
\label{obs_ptdouble}
\normalfont
 Let  $ \pi $= $\pi_{1} $, ..., $ \pi_{i}=\pi_{1}-1 $, $ \pi_{i+1}=a $,..., $ \pi_{n} $. A double with prefix transposition is possible iff a-1$\in [\pi_{1}...\pi_{i-1}  ) $.
 \end{obs}
$ \textbf{Proof :} $ $ (\rightarrow) $ Clearly( $\pi_{0} $, $ \pi_{1} $, ....., a-1),.....,$ \pi_{i}= \pi_{1-1},* \pi_{i+1}=a ,....\pi_{n-1} $is a double.
\\$ (\leftarrow) $To execute a double we must create an adjacency with the left end of the moved prefix $ [\pi_{0},....x] $ i.e. $ \pi_{0} $. Therefore the prefix is moved to a position just after $ \pi_{0} -1$. In order to create an adjacency at the right end of the moved prefix, $ x= \pi_{i+1}-1 $, i.e.  $ x=a-1 $. Further, $ \textit{x} $ is to the left of $ \pi  _{i} $ therefore
$ a-1 \in (\pi_{1}....\pi_{i-1}) $ $~\blacksquare  $ \\
A permutation that has no adjacencies is said to be irreducible. Likewise, if $ \pi $ can be reduced then the length of the resulting permutation $ \pi^{*} $ that cannot be reduced any further is called the reduced length of $ \pi $. For n$ \geq $K,  P$_{n}(k) $
denotes the subset of P$_{n} $ where the reduced length of any $ \pi$ $\in$ P$_{n}(k) $ is K.\\
The following theorem evaluates the expected probability that a $ \pi$ $\in$ P$_{n}(n) $ admits a double. In a reduced permutation also a given symbol is equally likely in all positions. For n=3, i.e. for the set $\{$(0 2 1), (1 0 2),(2 1 0)$\}$ this assumption holds. For n=4, i.e. for the set $ \{$ (0 2 1 3), (0 2 3 1), (0 3 2 1), (1 0 3 2), (1 3 0 2),
(1 3 2 0), (2 0 3 1), (2 1 0 3), (2 1 3 0), ( 3 0 2 1), (3 1 0 2), (3 2 1 0)$\}$ this holds.\\
%
%
%
%
%
%
%
\begin{theorem}
\normalfont
\label{th_prob_double}
 The expected probability of  $ \pi $= ( $\pi_{1}$=f, $\pi_{2} $, ...$ \pi_{i}$=f-1, $\pi_{i+1}=a $,..., $ \pi_{n} $) $\in$ P$_n(0)$ admitting a double is $ \sigma $ = $ \frac{1}{2}$ $-$  $\frac{2}{n(n-1)}$.
 \end{theorem}
$ \textbf{Proof :} $ General form of $ \pi $ is $f,\ldots, f-1, a\ldots$ where $f=  \pi_{1} $ and $a$ is the element succeeding $f-1$. Let $p$ be the probability of a double. Any permutation where $f=  \pi_{1} $ and $f-1 \in$ an arbitrary position in $ [2,...,n] $ occurs with a probability of $\frac{1}{n(n-1)}  $.
Here $f= \pi_{1} $  with a probability of $ \frac{1}{n} $ and $f-1$ can be in any of the positions $ [2,...,n] $ with an equal probability of $ \frac{1}{n-1} $.
We analyze the cases where $f=0,\ldots,n-1$. In each case the position of $f-1$ forms a sub-case. Here we denote the probability of a double where $f=k$ (or $f \in S$, a set of symbols) as $ \sigma (k)$ (resp. $ \sigma $(S)). Likewise, $ \sigma(k, j)$ ($ \sigma( S, j )$) is the probability of a double where $f=k$ (resp. $ \in S$ ) and the position of $f-1$ is $j$.\\
\textbf{Case(i):} $f=0$. $ \sigma $(0)=0. Here the moved prefix cannot create a new adjacency at the left end. Thus, a double is not possible.\\
\textbf{Case(ii)}: $f=n-1$. $ \sigma $(n-1) =$\frac{(n-3)}{2n(n-1)}$  + $\frac{1}{n(n-1)}$. 
\\The following subcases are partitioned as per the index of n-2, i.e. $ \pi^{-1} _{n-2} $. Note that $ \pi_{n-2}  $ is the element at the position n-2. Thus, the total probability of $f, f-1$ and $a$ being in their respective positions is $ \mu= $ $\frac{1}{n(n-1)(n-2)}  $
\begin{itemize}
\item $ \sigma(n-1,2)=0 $. The moved prefix i.e. n-1 cannot create an adjacency at the right end. That is $(n-1)+1 \notin \Sigma $.
\end{itemize}
\begin{itemize}
\item $ \sigma(n-1,3)=\frac{\mu}{n-3} $. Here one can execute a double if $ \pi_{2}= a-1$. Here for $a =1,\ldots,n-3$, $\pi_{2} = a-1 $ with a probability of $ \frac{1}{n-3} $. Note that out of n-3 positions only one position is favorable.
\end{itemize}
\begin{itemize}
\item $ \sigma(n-1,4)=\frac{2\mu}{n-3} $. Here one can execute a double if $ \pi_{2}= a-1$ or $ \pi_{3} = a-1$. Here for $a =1,\ldots,n-3$, $ \pi_{2} $ = a-1 with a probability of $ \frac{2}{n-3} $. Thus out of n-3 positions only two position are favorable.
\end{itemize}
\begin{itemize}
\item $ \sigma(n-1,n-1)=\frac{(n-3)\mu}{n-3} $. Here one can execute a double if $ \pi_{2} = a-1$ or $ \pi_{3}= a-1 \ldots \pi_{n-2}= a-1$. Here for $a =1,\ldots,n-3$, one of $ \pi_{2} $ ,...$ \pi_{n-2}= a-1$ with a probability of $ \frac{n-3}{n-3} $.\\
Thus the total probability for a particular value of a is the summation of the above probabilities. That is, $0+\frac{1}{n(n-1)(n-2)}$ $( $ $\frac{1}{(n-3)}  $ +$\frac{2}{(n-3)}$ +$\frac{3}{(n-3)}$ +.......+ $\frac{n-3}{(n-3)}$ $ )$$ =\frac{1}{2n(n-1)} $
\end{itemize}
\begin{itemize}
\item Here $a$ can take values from 1, 2, 3,......, n-3 (a cannot be f or f-1 or 0. a=0 is not possible since $a-1 \notin\Sigma $). Thus, $n-3$ values are possible for $a$. So,  $ \sigma $(n-1,[2 \ldots n-1]) =$\frac{(n-3)}{2n(n-1)}$.
 \end{itemize}
\begin{itemize}
\item Type 2 adjacency includes an adjacency between $n-1$ in the last position and (imagined) $\pi_{n+1} =n$ .
Thus, if $f-1$ occurs in the last position then one can execute $[f, \ldots n-1] \ldots f-1 *$ and create two new adjacencies, one between f-1 and f and another one between n-1 and imagined n. Thus, $ \sigma $(n-1,n) =$\frac{1}{n(n-1)}$.
 \end{itemize}
%
%
\textbf{Case(iii)}:  $f \neq $n-1 and $f\neq $0; i.e. $f \in S$ where $S=\{1,2,3,....,n-2 \} $.
 $ \sigma (S) =\frac{1}{2}-\frac{3}{2(n-1)} - \frac{1}{2(n-1)(n-2)}- \frac{1}{n(n-1)(n-2)} + \frac{n-2}{n(n-1)}$.

 This case is partitioned into two sub cases. Case (iii-a) $f=1$. Case (iii-b) $f > 1$. Similar to Case(ii) the probability of $f$ and $f-1$ to be positioned in their respective positions equals $ \frac{1}{n(n-1)} $. Also, $a$ occurs in its position with $\frac{1}{(n-2)}$ probability.\\
  %
  %
 $ \textbf{ Case (iii-a)} $: Here $f=1$. $ \sigma(1)= \frac{(n-3)}{2(n-1)(n-2)} - \frac{1}{n(n-1)(n-2)} $ + $\frac{1}{n(n-1)}$.\\
Consider $f=1, f-1=0 = \pi_{2} $ and $a =f+1=2$. Here one can move $f$ in between $f-1$ and $a$. Further, the same move works if $f-1$ is any of the positions $3 \ldots n-1$). Given these configurations the probability of a double  is 1. So, the probability when $a=f+1$ is  $ \frac{(n-2)}{n(n-1)(n-2)} $=$ \frac{1}{n(n-1)} $. $ (A)$\\
$ \hspace*{1cm} $ If $0 = \pi_{3}(\pi_{k}) $  then for a double $a-1=  \pi_{2}(\pi_{<k}) $. So, for a particular value of $a$ the probability is $\frac{1}{n(n-1)(n-2)}$ $( $ $\frac{1}{(n-3)}  $ +$\frac{2}{(n-3)}$ +$\frac{3}{(n-3)}$ +.......+ $=\frac{n-3}{(n-3)}$ $ )$$ =\frac{1}{2n(n-1)} $\\
 Here $a \in [4, 5,\ldots, n-1]$ ($a \notin \{0, 1, f+1=2, f+2=3\}$;  $f+1=2$ is analyzed above and $f+2=3$ is  analyzed below. So for all the $n-4$ values the combined probability is $ \frac{(n-4)}{2n(n-1)} $. $(B) $\\
$ \hspace*{1cm} $ When $a=f+2=3$ and $ f-1=0=\pi_{3} $ a double is infeasible because $ \pi $ = (1 2 0 3.....) is disallowed due to the presence of the adjacency $ \textbf{1 2} $. Recall that we only consider the reduced permutations.  When $f-1$ comes in the fourth position then for a double to be feasible $ a-1=2$ can be in the third position. So, given $f$ and $f-1$, the probability of a double is$ \frac{1}{n-3} $. Similarly the probabilities for the other positions of $f-1$ are deduced. Recall that $a-1 \neq  \pi_{2} $. So, the total probability for this subcase is\\
  $\frac{1}{n(n-1)(n-2)}$ $( $ $\frac{1}{(n-3)}  $ +$\frac{2}{(n-3)}$ +$\frac{3}{(n-3)}$ +.......+ $\frac{n-4}{(n-3)}$ $ )$$ =\frac{(n-4)}{2n(n-1)(n-2)} $ $(C) $.\\\\
  Thus, the total probability of case(iii-a) $ \sigma(1)=A+B+C-K$ where $K$ is the probability corresponding to $ \pi_{n}= a=n-1$. Note that in Type 2 adjacency $ \pi_{n}=n-1$ creates an adjacency and we assume that the permutation is reduced, so, this scenario is not possible. Here, $K=\frac{1}{n(n-1)(n-2)}$.
  The total probability of case(iii-a) equals $ \sigma (1)=A+B+C-K=$\\\\
  $ \frac{1}{n(n-1)} + \frac{(n-4)}{2n(n-1)} + \frac{(n-4)}{2n(n-1)(n-2)} -$ $\frac{1}{n(n-1)(n-2)} = \frac{2(n-2)+(n-4)(n-2)+(n-4)}{2n(n-1)(n-2)} - \frac{1}{n(n-1)(n-2)} $\\\\
 $= \frac{2n-4+n^{2}-6n+8+(n-4)}{2n(n-1)(n-2)} - \frac{1}{n(n-1)(n-2)} $
  $=\frac{(n-3)}{2(n-1)(n-2)} - \frac{1}{n(n-1)(n-2)} $ \\
 \begin{itemize}
 \item Type 2 adjacency includes an adjacency between $\pi_{n}= n-1$ and imagined $\pi_{n+1}= n$.  Thus, if $f-1$ occurs in the last position then one can execute $[f, \ldots n-1] \ldots f-1 *] $ and create two new adjacencies, one between $f-1$ and $f$ and another one between $n-1$ and imagined $n$ in position number $n+1$. Thus, $ \sigma $(1,n) =$\frac{1}{n(n-1)}$. When this probability is also included then the total probability for Case (iii-a) is:
 $\frac{(n-3)}{2(n-1)(n-2)} $ - $ \frac{1}{n(n-1)(n-2)} $ + $\frac{1}{n(n-1)}$ 
  \end{itemize}
 %
 %
 %
 %
 \textbf{Case(iii-b):} $f> $ 1  i.e. $S= \{2,3,....,n-2 \} $.
  $ \sigma $(S) =$\frac{(n-3)(n^{2}-4n)}{2n(n-1)(n-2)}$  + $\frac{n-3}{n(n-1)}$. \\
  \\The probability that $f=\pi_{1} $ and $f-1= \pi_{j} $ (for $j \neq $1) is $ \frac{1}{n(n-1)} $. The probability that $a= \pi_{j+1} $ is $ \frac{1}{n-2} $. When $f-1$ immediately succeeds $f$ then $a$ must be $f+1$. When $ a=f+1$,  $f-1$ can be in any of the positions $[2 \ldots n-1]$. Given this configuration the probability of a double is 1. So the total probability corresponding to $f-1$ occupying all possible positions is $ \frac{(n-2)}{n(n-1)(n-2)} $= $ \frac{1}{n(n-1)}$. $(P) $\\
  $ \hspace*{1cm} $ If $f-1 =\pi_{3} $ then for a double to be feasible $a-1= \pi_{2}$.
 Given that  $f-1 =\pi_{3} $, $a-1= \pi_{2} $ with a probability of $ \frac{1}{n-3} $. The denominator is $n-3$ because we exclude $f, f-1$ and $a$. Similarly, for $f-1 =( \pi_{4} $ \ldots $ \pi_{n-1} $) we obtain the respective probabilities as ($ \frac{2}{n-3} $,$ \frac{3}{n-3} $, ...$ \frac{(n-3)}{(n-3)} $ ).
 Thus, for particular values of $f$ and $a$ the probability of a  double in this case is  $=\frac{1}{n(n-1)(n-2)}$ $( $ $\frac{1}{(n-3)}  $ +$\frac{2}{(n-3)}$ +$\frac{3}{(n-3)}$ +.......+ $\frac{n-3}{(n-3)}$ $ )$$ =\frac{1}{2n(n-1)}. $\\\\
 Here $a$ can take $n-5$ values which exclude $0, f, f-1,a= f+1$ and $a=f+2$ where $a= f+1$ is considered above and $a=f+2$ is analyzed below. 
 Thus, the probability for all values of $a$ is $ \frac{(n-5)}{2n(n-1)}$. $(Q) $\\\\
 $ \hspace*{1cm} $ When $a=f+2$ and $f-1 = \pi_{3}$ then double is infeasible because $\pi = ( f, f+1, f-1 a=f+2,\ldots.)$ is not reduced. If $f-1$ is in the fourth position then for a double to occur $a-1$ can be in the third position only. So the probability of $a-1$ occurring  between $f$ and $f-1$ is $ \frac{1}{(n-3)}$ (we exclude $f, f-1$ and $a$). 
 Similarly, the probabilities of a double are derived when $f-1$ occurs in other positions. If $f-$1 is in the$(n-1)^{th}$, position then for a  double to be feasible $a-1$ can reside in any of the $(n-4)$ positions. The probability of $a-1$ residing between $f$ and $f-1$ is  $\frac{(n-4)}{(n-3)}$. So, the total probability for this case =	$\frac{1}{n(n-1)(n-2)}$ $( $ $\frac{1}{(n-3)}  $ +$\frac{2}{(n-3)}$ +$\frac{3}{(n-3)}$ +.......+ $\frac{n-4}{(n-3)}$ $ )$
 $ =\frac{(n-4)}{2n(n-1)(n-2)} $ ($R$) \\
 
  Recall that K is the probability corresponding to $ \pi_{n}= a=n-1$. As the permutation is reduced this scenario is avoided. Here $f$ can assume $n-3$ values that remain after excluding $0, n-1$, and $1$.  
 Total probability for a particular value of $f$ in Case(iii-b) equals $P+Q +R-K$ =\\\\
 $ \frac{1}{n(n-1)} $+$ \frac{(n-5)}{2n(n-1)} $+$ \frac{(n-4)}{2n(n-1)(n-2)}  -K$= \\\\
 $ \frac{2(n-2)+(n-5)(n-2)+(n-4)}{2n(n-1)(n-2)} - \frac{1}{n(n-1)(n-2)} $=\\\\
  $\frac{2n-4+n^{2}-7n+10+(n-4)}{2n(n-1)(n-2)} -\frac{1}{n(n-1)(n-2)} $=$\frac{n^{2}-4n+2 -2}{2(n-1)(n-2)} $  $= \frac{n^{2}-4n}{2(n-1)(n-2)}$.\\
 
\begin{itemize}
 \item Type 2 adjacency includes an adjacency between $\pi_{n}= n-1$ and the imagined $\pi_{n+1}= n$. Thus, if $f-1$ occurs in the last position then one can execute $[f, \ldots n-1] \ldots f-1 *$ and create two new adjacencies, one between $f-1$ and $f$ and another one between $n-1$ and imagined $ n$. There are $n-3$  choices for $f$, so, $\sigma([2 \ldots n-2],n)$ =$\frac{n-3}{n(n-1)}$.
\end{itemize}
    
  So, the total probability for all values of $f$ for   Case(iii-b) =$\frac{(n-3)(n^{2}-4n)}{2n(n-1)(n-2)}+$	$\frac{n-3}{n(n-1)}$. 
  The total probability for Case(iii) that is partitioned into Case(iii-a) and Case(iii-b) is therefore\\\\
$\frac{(n-3)}{2(n-1)(n-2)}- \frac{1}{n(n-1)(n-2)}$ + $\frac{(n-3)(n^{2}-4n)}{2n(n-1)(n-2)}$ + 	$\frac{n-3+1}{n(n-1)}$\\\\
$= \frac{(n-3)}{2n(n-1)(n-2)}$$ (n^{2}-4n+n) - K + \frac{n-3+1}{n(n-1)}$\\
\\=$\frac{(n-3)}{2n(n-1)(n-2)}$($n^{2}-3n) -K + \frac{n-3+1}{n(n-1)}=$\\\\
$\frac{(n-3)(n-3)}{2(n-1)(n-2)}-\frac{1}{n(n-1)(n-2)} $ + $\frac{n-3+1}{n(n-1)}=$\\\\
$ \frac{1}{2}  -  \frac{3}{2(n-1)}  + \frac{1}{2(n-2)(n-1)} - \frac{1}{n(n-1)(n-2)} $ + $\frac{n-3+1}{n(n-1)}$.
We add the probability of case(ii) to this expression to obtain the final probability as:\\
$\frac{1}{2} - $ $ \frac{2}{n(n-1)}$.

Thus, the total probability for all cases that is $\sigma(\Sigma)$ = $\frac{1}{2} - $ $ \frac{2}{n(n-1)}$.$~\blacksquare$	

\begin{obs}
\label{obs_ptexpadj}
\normalfont
Let $ \pi\in $P$ _{n} $ be reduced then the expected number of new adjacencies created per move is 1+$ \sigma $.
\end{obs}
$ \textbf{Proof}: $ A double can be executed with a probability of $ \sigma $ and it creates two new adjacencies and a single can be executed with a probability of (1-$ \sigma $) and it creates one new adjacency. Thus, in a move the
expected number of new adjacencies created= $ \sigma $(2) + (1-$ \sigma $)(1)= 1+$ \sigma $. Note that from Theorem \ref{th_prob_double}, for prefix transpositions, this measure $\approx 1.5$ for large values of n.$~\blacksquare$


\section{Results and Conclusions}
The  average number of moves to sort all permutations in P$_n(0)$ for $n \eqslantless 9$ and the average number of moves to sort all permutations in P$_n$ for $ 1 < n \eqslantless 9$ are computed and shown in Tables 3 and 4. When n=0, no moves are needed.
In each of these table the computed values of P$_j(0),~j \eqslantless i$ are used as a basis to predict the rest of the values. The predicted values can be computed for large values of n because given the computed values of P$_j(0),~j \eqslantless i$ the model runs in O($n^2$). The predicted values can be compared to the computed values for $n \eqslantless 9$. Currently, for $n>9$ the computed values are not applicable, denoted by: na.
For n = $2 \ldots i$ the values are computed thus the predicted values are shown as dashes.  The model gets more accurate for larger values of n.

We compute $\forall {k} |P_n(k)|$ in O(n$ ^{2} $) time. We show that the number of irreducible permutations in $P_n$ is $\Theta(n!)$. The computation of $\forall {k} |P_n(k)|$ leads to a framework for analysis of  block-moves on permutations. Bulteau et al. show that sorting permutations by transpositions is NP-hard \cite {bulteau}. Thus, it is desirable to compute the expected number of moves to sort a permutation $ \in P _{n} $ with transpositions. It is believed that sorting permutations by either prefix transpositions or suffix transpositions or prefix/suffix transpositions is also NP-hard. So, the computation of say  $ E(P _{n}) $, the expected number of moves to sort permutations,  with various block-moves is of interest. $ E(P _{n}) $ for a particular operation indicates the expected number of moves a packet from some source node $u$ to some destination node $v$  must traverse in the corresponding Cayley graph. 

  The main contribution of this article is the theoretical framework for estimating the expected number of moves, i.e. $E(P_{n})$ to sort permutations in P$_n$ with various block-moves  in O($n^2$) time; given the computation of $E(P_{i})$ for some $i<n$.  We employ a model based on the proposed framework to estimate $E(P_{n})$ in O($n^2$) time for prefix transpositions. Due to symmetry, the corresponding  results are applicable for suffix transpositions also. Based on this model, Pai and Kumarasamy worked on estimating $E(P_{n})$ with transpositions \cite{paiK15}. Current work is focused on exploring models for estimation.

{\small
}

{\footnotesize
\bibliographystyle{abbrv}
\bibliography{Adj_Perm}
}

\begin{table}[!htbp]
{\footnotesize Table 0. Values of $\mid P_{n}(k)\mid $ for type 1 adjacency.}
 \tiny 
  {\fontsize{3.5}{5}\selectfont 
\begin{center}
     \begin{tabular}{ |c|c|c|c|c|c|c|c|c|c|c|c|c|c|c|c|c|c|} 
     n$ \setminus $k  & 0  & 1 & 2  & 3 & 4 & 5 & 6 & 7  & 8 & 9 & 10 & 11  & 12 & 13 & 14 & $ \Sigma\mid P_{n}(k)\mid $ \\
     
      \hline
       2  & 1  & 1 & 0  & 0 & 0 & 0 & 0 & 0  & 0 & 0 & 0 & 0  & 0 & 0 & 0 & 2 \\
           
      3  & 3  & 2 & 1  & 0 & 0 & 0 & 0 & 0  & 0 & 0 & 0 & 0  & 0 & 0 & 0 & 6 \\
                      
      4  & 11  & 9 & 3  & 1 & 0 & 0 & 0 & 0  & 0 & 0 & 0 & 0  & 0 & 0 & 0 & 24 \\
                            
     5  & 53  & 44 & 18  & 4 & 1 & 0 & 0 & 0  & 0 & 0 & 0 & 0  & 0 & 0 & 0 & 120 \\
     6  & 309  & 265 & 110  & 30 & 5 & 1 & 0 & 0  & 0 & 0 & 0 & 0  & 0 & 0 & 0 & 720 \\
     7 & 2119 & 1854 & 795 & 220 & 45 & 6 & 1 & 0 & 0& 0& 0& 0& 0& 0& 0 & 5040  \\
     8 &  16687& 14833& 6489& 1855& 385& 63& 7 & 1 & 0 & 0& 0& 0& 0& 0 & 0 & 40320 \\
     9 & 148329& 133496& 59332& 17304& 3710& 616 & 84& 8& 1& 0& 0& 0& 0& 0& 0&  362880   \\
     10& 1468457& 1334961& 600732& 177996& 38934& 6678& 924& 108& 9& 1& 0& 0& 0& 0& 0& 3628800\\                               
     11& 16019531& 14684570& 6674805& 2002440& 444990& 77868& 11130& 1320& 135& 10& 1& 0& 0& 0& 0& 39916800\\
     12& 190899411& 176214841& 80765135& 24474285& 5506710& 978978& 142758& 17490& 1815& 165& 11& 1& 0& 0& 0& 479001600\\
     13& 2467007773& 2290792932& 1057289046& 323060540& 73422855& 13216104& 1957956& 244728& 26235& 2420& 198& 12& 1& 0& 0& 6227020800\\
     14& 34361893981& 32071101049& 14890154058& 4581585866& 1049946755& 190899423& 28634892& 3636204& 397683& 37895& 3146& 234& 13& 1& 0& 87178291200                         
     \end{tabular}
     \end{center}
     }

\end{table}
\begin{table}[!htbp]
{\footnotesize Table 1. Values of $\mid P_{n}(k)\mid $ for the b-adjacency. These numbers hold for f-adjacency as well.}
 \tiny 
  {\fontsize{3.5}{5}\selectfont 
\begin{center}
     \begin{tabular}{ |c|c|c|c|c|c|c|c|c|c|c|c|c|c|c|c|c|c|} 
     n$ \setminus $k  & 0  & 1 & 2  & 3 & 4 & 5 & 6 & 7  & 8 & 9 & 10 & 11  & 12 & 13 & 14 & $ \Sigma\mid P_{n}(k)\mid $ \\
     
      \hline
       2  & 1  & 0 & 1  & 0 & 0 & 0 & 0 & 0  & 0 & 0 & 0 & 0  & 0 & 0 & 0 & 2 \\
           
      3  & 2  & 3 & 0  & 1 & 0 & 0 & 0 & 0  & 0 & 0 & 0 & 0  & 0 & 0 & 0 & 6 \\
                      
      4  & 9  & 8 & 6  & 0 & 1 & 0 & 0 & 0  & 0 & 0 & 0 & 0  & 0 & 0 & 0 & 24 \\
                            
     5  & 44  & 45 & 20  & 10 & 0 & 1 & 0 & 0  & 0 & 0 & 0 & 0  & 0 & 0 & 0 & 120 \\
     6  & 265  & 264 & 135  & 40 & 15 & 0 & 1 & 0  & 0 & 0 & 0 & 0  & 0 & 0 & 0 & 720 \\
     7 & 1854 & 1855 & 924 & 315 & 70 & 21 & 0 & 1 & 0& 0& 0& 0& 0& 0& 0 & 5040  \\
     8& 14833& 14832& 7420& 2464& 630& 112& 28& 0& 1& 0& 0& 0& 0& 0& 0& 40320\\
     9& 133496& 133497& 66744& 22260& 5544& 1134& 168& 36& 0& 1& 0& 0& 0& 0& 0& 362880\\
     10& 1334961& 1334960& 667485& 222480& 55650& 11088& 1890& 240& 45& 0& 1& 0& 0& 0& 0& 3628800\\
     11& 14684570& 14684571& 7342280& 2447445& 611820& 122430& 20328& 2970 & 330& 55& 0& 1& 0& 0& 0& 39916800\\
     12& 176214841& 176214840& 88107426& 29369120& 7342335& 1468368& 244860& 34848& 4455& 440& 66& 0& 1& 0& 0& 479001600 \\
     13& 2290792932& 2290792933& 1145396460& 381798846& 95449640& 19090071& 3181464& 454740& 56628& 6435& 572& 78& 0& 1& 0& 6227020800 \\
     14& 32071101049& 32071101048& 16035550531& 5345183480& 1336295961& 267258992& 44543499& 6362928& 795795& 88088& 9009& 728& 91& 0& 1& 87178291200                    
     \end{tabular}
     \end{center}
     }

\end{table}
\begin{table}[!htbp]
{\footnotesize Table 2. Values of $\mid P_{n}(k)\mid $ for the  bf-adjacency.}
 \tiny 
  {\fontsize{3}{5}\selectfont 
\begin{center}
     \begin{tabular}{ |c|c|c|c|c|c|c|c|c|c|c|c|c|c|c|c|c|c|c} 
     n$ \setminus $k  & 0  & 1 & 2  & 3 & 4 & 5 & 6 & 7  & 8 & 9 & 10 & 11  & 12 & 13 & 14 & 15& $ \Sigma\mid P_{n}(k)\mid $ \\
     
      \hline
       2  & 1  & 0 & 0  & 1 & 0 & 0 & 0 & 0  & 0 & 0 & 0 & 0  & 0 & 0 & 0 & 0& 2 \\
           
      3  & 1  & 4 & 0  & 0 & 1 & 0 & 0 & 0  & 0 & 0 & 0 & 0  & 0 & 0 & 0 &0& 6 \\
                      
      4  & 8  & 5 & 10  & 0 & 0 & 1 & 0 & 0  & 0 & 0 & 0 & 0  & 0 & 0 & 0 & 0& 24 \\
                            
     5  & 36  & 48 & 15  & 20 & 0 & 0 & 1 & 0  & 0 & 0 & 0 & 0  & 0 & 0 & 0 & 0& 120 \\
     6& 229& 252& 168& 35& 35& 0& 0& 1& 0& 0& 0& 0& 0& 0& 0& 0& 720 \\
     7& 1625& 1832& 1008& 448& 70& 56& 0& 0& 1& 0& 0& 0& 0& 0& 0& 0& 5040  \\
     8& 13208& 14625& 8244& 3024& 1008& 126& 84& 0& 0& 1& 0& 0& 0& 0& 0& 0& 40320 \\
     9& 120288& 132080& 73125& 27480& 7560& 2016& 210& 120& 0& 0& 1& 0& 0& 0& 0& 0& 362880 \\
     10& 1214673& 1323168& 726440& 268125& 75570& 16632& 3696& 330& 165& 0& 0& 1& 0& 0& 0& 0& 3628800 \\
      11& 13469897& 14576076& 7939008& 2905760& 804375& 181368& 33264& 6336& 495& 220& 0& 0& 1& 0& 0& 0& 39916800 \\
      12& 162744944& 175108661& 94744494& 34402368& 9443720& 2091375& 392964& 61776& 10296& 715& 286& 0& 0& 1& 0& 0& 479001600 \\
      13& 2128047988& 2278429216& 1225760627& 442140972& 120408288& 26442416& 4879875& 785928& 108108& 16016& 1001& 364& 0& 0& 1& 0& 6227020800 \\
      14& 29943053061& 31920719820& 17088219120& 6128803135& 1658028645& 361224864& 66106040& 10456875& 1473615& 180180& 24024& 1365& 455& 0& 0& 1& 87178291200         
     \end{tabular}
     \end{center}
     }

\end{table}

\begin{table}[!htbp]
{\footnotesize Table 3. Computed and Predicted values of P$_{n}$(0). Initialization: P$_{1}$(0)...P$_{i}$(0). 
\begin{center}
\begin{tabular}{ |c|c|c|c|c|c|c|c|c|c|c|c|c|c|c|c|c|c}
\hline
n  & 2 & 3 & 4 & 5 & 6 & 7 & 8 & 9 & 10 & 11 & 12 & 13& 14 & 15 & 16   \\
\hline
Computed &1.0 &2.0 &2.33 &3.09&3.68&4.29&4.91&5.50&na &na &na &na &na &na &na \\
\hline
Pred. i=6 &- &- &- &- &- &4.21 &4.81 &5.43 &6.07 &6.71 &7.37 &8.02 &8.69 &9.35 &10.01\\
Pred. i=7 &- &- &- &- &- &- &4.83 &5.46 &6.08 &6.72 &7.37 &8.03 &8.69 &9.35 &10.01\\
Pred. i=8 &- &- &- &- &- &- &- &5.47 &6.10 &6.73 &7.38 &8,03 &8.69 &9.35 &10.01\\
\hline
\end{tabular}
\end{center}
}
\end{table}

\begin{table}[!htbp]
{\footnotesize Table 4. Computed and Predicted values of E(X$ _{n}$). Initialization: P$_{1}$(0)...P$_{i}$(0)
\begin{center}
\begin{tabular}{ |c|c|c|c|c|c|c|c|c|c|c|c|c|c|c|c|c|c} 
\hline
n  & 2 & 3 & 4 & 5 & 6 & 7 & 8 & 9 & 10 & 11 & 12 & 13& 14 & 15 & 16   \\
\hline
Computed &0.5 &1.16 &1.79 &2.42 &3.06 &3.68 &4.29 &4.90 &na &na &na &na &na &na &na \\ 
\hline
Pred. i=6 &- &- &- &- &- &3.65 &4.23 &4.82 &5.44 &6.07 &6.72 &7.37 &8.03 &8.69 &9.35 \\ 
Pred. i=7 &- &- &- &- &- &- &4.26 &4.86 &5.46 &6.09 &6.73 &7.38 &8.03 &8.69 &9.35 \\ 
Pred. i=8 &- &- &- &- &- &- &- &4.89 &5.50 &6.11 &6.74 &7.38 &8.03 &8.69 &9.35 \\ 
\hline
\end{tabular}
\end{center}
}
\end{table}

\section {Acknowledgments}
 This article studies adjacencies in permutations as defined in \cite{chitturi08}. It is based on the tech. report UTDCS-03-15, dept. of CS, Univ. of Texas at Dallas. The articles \cite{Roselle, whitworth} and the existence of integer sequences  equivalent to the ones studied in the current article came to our attention due to an anonymous referee.

\section {Appendix}

The proofs for Theorem 1-3  are deferred to the appendix. This is intended to improve the readability of the main text. The proof of Theorem 2 is similar to that of Theorem 3 and hence it is omitted.
\\
 \textbf{Theorem 1} Let P$ _{n} $(k) be the subset of P$ _{n} $ where any $ \pi $$ \in $ P$ _{n} $(k) has exactly k type 1 adjacencies. Let f(n,k) be the cardinality of P$ _{n} $(k). Then f(n,k) = f(n$ - $1,k$ - $1)+ (n$ - $1$ - $k) * f(n$ - $1,k) + (k+1)* f(n$ - $1,k+1) where $ 0 \leq $k$<$n.
\\
$ \textbf{Proof~} $ 
We denote the number of adjacencies in a permutation $ \pi $$ \in $P$ _{n} $ with $ \alpha $($ \pi $) and the number of permutations $ \in $P$ _{n} $ having $ \alpha $($ \pi $) adjacencies with f(n, $ \alpha $($ \pi $)). Recall that $ \Sigma $ =$\{$0,....,n$ - $1 $\}$. Thus, $ \pi^{*} $$ \in $P$ _{n-1} $ is
composed of $\{$0,....,n$-$2$\}$and a member of P$ _{n} $ additionally contains n$ -$1. 
Let $ \alpha $($ \pi^{*} $)=q for $ \pi^{*} $$ \in $P$ _{n-1} $.
When a $ \pi $$ \in $P$ _{n} $ is formed from $ \pi^{*} $ by inserting n$ - $1 we have the following three cases: (i) $ \alpha $($ \pi $)=$ \alpha $($ \pi^{*} $), (ii)
$ \alpha $($ \pi $)=$ \alpha $($ \pi^{*} $)+1 and (iii) $ \alpha $($ \pi $)=$ \alpha $($ \pi^{*} $)$-$1.\\
\hspace*{1.5cm} The element n$-$1 can create an adjacency only if it immediately succeeds n$ - $2; however in
such a case it cannot destroy an existing adjacency; i.e. $ \alpha $($ \pi $)=$ \alpha $($ \pi^{*} $)+1 (Case (ii)). 
Thus, it is not possible to insert n$-$1 in any position where it simultaneously creates one adjacency and destroys one adjacency.
Similarly, if n$-$1 is inserted between a, a+1 for some a then $ \alpha $($ \pi $)=$ \alpha $($ \pi^{*} $)$-$1 (Case (iii)). Finally, if n$-$1 neither succeeds n$-$2 nor splits a, a+1 (for some a) then $ \alpha $($ \pi $)=$ \alpha $($ \pi^{*} $) (Case (i)).
 We determine the number of permutations $ \pi $ of P$ _{n} $ with $ \alpha $($ \pi $)=k that can be generated from some permutations in P$ _{n-1} $ corresponding to each of these cases.
 
\hspace*{1.5cm}Let $  $($ \pi^{*} $)= k. For a given $ \pi^{*} $ we want to determine  how many $ \pi $$ \in $P$ _{n} $ exist such that $ \alpha $($ \pi^{*} $)=$ \alpha $($ \pi$). In order to generate $ \pi$$ \in $P$ _{n} $ from $ \pi^{*} $ $\in$ P$ _{n-1} $ one can insert n$ - $1 in any of the n positions (n$ - $2 internal and two
extreme positions). However, in order to maintain the same number of adjacencies, k+1 positions are
forbidden; where k positions correspond to existing adjacencies and one corresponds to the adjacency that can be created between n$ - $1 and n. Thus, the contribution of f(n$ - $1,k) to f(n,k) is (n$ - $1$ - $k) * f(n$ - $1,k). Let $ \alpha $($ \pi^{*} $)= k$ - $1, we want to determine the contribution of f(n$ - $1,k) to f(n,k). In order to create a $ \pi $$ \in $P$ _{n} $ from $ \pi^{*} $ where $ \alpha $($ \pi $)= k the only possibility is that n$ - $1 is inserted to the immediate right of n$ - $2. Thus, f(n$ - $1,k) contributes to f(n,k) exactly f(n$ - $1,k). Finally, we want to determine the contribution of f(n$ - $1,k+1)
to f(n,k). Here, any one of the k+1 adjacencies can be broken by inserting n+1 in between. Thus, the contribution of f(n$ - $1,k+1) to f(n,k) is (k+1)* f(n$ - $1,k+1). Note, that f(n,k) is restricted to the above cases. The corresponding  algorithm, Adjacency\_Count, is given below. The theorem follows.$~\blacksquare$  
\\
{\footnotesize
\begin{algorithmic}
\State Algorithm: $ \textbf{Adjacency\_Count} $

\State Initialization: f(2,0)=1;f(2,1)=1;$ \forall_{i} $f(i,i$-$1)=1
\For{(i=3,....n)}
\For{(k=0,....i$-$2)}
 \If{(k==0) }f(i,k) = (i$-$1) * f(i$-$1,0) + f(i$-$1,1);
 \Else 
 \If{(k==i$-$2)} f(i,k) = f(i$-$1,k$-$1) + f(i$-$1,k); \Else f(i,k) = f(i$ - $1,k$ - $1)+ (i$ - $k$ - $1) * f(i$ - $1,k) + (k+1)* f(i$ - $1,k+1); \EndIf
 \EndIf
\EndFor
 \EndFor
\end{algorithmic}
}
%
\vspace{8pt}
\noindent Theorem 3 is restated here. The proof accompanies. \\\\
\textbf{Theorem 3} Let f(i, j) denote the number of permutations in P$ _{i} $ with exactly j adjacencies. Then the
recurrence relation for f(i, j) is:\\
f(i, j)= (f(i$-$1, j)$-$f(i$-$2, j$-$1))*(i$-$j$-$2) +\\
(f(i$-$1, j$-$1)$-$f(i$-$2, j$-$2))*2 +f(i$-$2, j$-$2) +\\
(f(i$-$1, j+1)$-$f(i$-$2, j))*(j+1)+f(i$-$2, j)*(i$-$j$-$1)+\\
 f(i$-$2, j+1)*(j+1); 0$ \leq $ j$ \leq $i+1.
\\\\
$ \textbf{Proof~} $ We presume that $\pi_{n+1} $ =n and $\pi_{0}=-1 $. So, if $\pi_{n}$ =n$ - $1 then $\pi_{n}$ and $\pi_{n+1}$ form an adjacency.
Likewise, if $\pi_{1}=0$ then $\pi_{0}$ and $\pi_{1}$ form an adjacency.
We denote the number of adjacencies in $ \pi $$ \in $P$ _{n} $ with $ \alpha $($ \pi $) and the number of permutations$ \in $P$ _{n} $ having $ \alpha $($ \pi $) adjacencies
with f(n, $ \alpha $($ \pi $)). Recall that $ \Sigma $ =$\{$0, ....., n$-$1$\}$. Thus, $ \pi^{*} $$ \in $P$ _{n-1} $ is composed of $\{$0, ...., n$ - $2$\}$ and $ \pi $$ \in $P$ _{n} $
additionally contains n$ $ - $ $1. The formation of $ \pi $$ \in $P$ _{n} $ from $ \pi^{*} $ $ \in $P$ _{n-1} $ by inserting n$ - $1 can be partitioned into:
$ \textbf{Case (i)} $ $ \alpha $($\pi $)=$ \alpha $($ \pi^{*} $), $ \textbf{Case (ii)} $ $ \alpha $($\pi $)=$ \alpha $($ \pi^{*} $)+1, $ \textbf{Case (iii)} $ $ \alpha $($\pi $)=$ \alpha $($ \pi^{*} $)$-$1 and $ \textbf{Case (iv)} $ $ \alpha $($\pi $)=$ \alpha $($ \pi^{*} $)$-$2\\
\hspace*{1.5cm}The element n$ - $1 can create an additional adjacency only if it immediately succeeds n$ - $2 or $ \pi_{n} $ =n$ - $1.
However in such a case it cannot destroy an existing adjacency; i.e. $ \alpha $($ \pi $)=$ \alpha $($ \pi^{*} $)+1 \textbf{(Case (ii))}. Thus, it is not possible to insert n$ - $1 in any position where it simultaneously creates one adjacency and destroys one adjacency. 
Similarly, if n$ - $1 is inserted between x, x+1 for some x then $ \alpha $($ \pi $)=$ \alpha $($ \pi^{*} $)$ - $1 \textbf{(Case (iii))}. If n$ - $1 neither creates an adjacency nor splits any x, x+1 then $\alpha $($ \pi $)=$ \alpha $($ \pi^{*} $) \textbf{(Case (i))}. Consider $ \pi^{*} $ where $ \pi^{*}_{n-1} $= n$ - $2; here $ \pi^{*}_{n-1} $ and (imagined) $ \pi^{*}_{n} $ form an adjacency; thus, if n$ - $1 is inserted into $ \pi^{*} $ in a position other
than n then $ \alpha $($ \pi $)=$ \alpha $($ \pi^{*} $)$ - $1. Furthermore, if n$-$1 breaks an existing adjacency in $ \pi^{*} $ then $ \alpha $($ \pi $)=$ \alpha $($ \pi^{*} $)$ - $2 \textbf{(Case (iv))}.\\
\hspace*{1.5cm}Given $ \pi^{*} $ $ \in $P$ _{n-1} $ where $ \alpha ( \pi^{*} )=p$, we determine the number of permutations in P$ _{n} $ that are generated from $ \pi^{*} $ that have k adjacencies. From the above discussion $p \in $$\{$k$ - $1, k, k+1, k+2$\}$. First, we observe the following.\\
$\textit{Observation:}$ The number of permutations in P$ _{n-1} $ with k adjacencies where $ \pi^{*}_{n-1}$= n$ - $2 equals f( n$ - $2, k$ - $1). Justification: If we disregard the last element the remaining elements that belong to P$ _{n-2} $ need to form k$ - $1 adjacencies. The observation follows.\\
$ \textbf{Case(i):} $ p=k. Here we determine the number of permutations in P$ _{n} $ that $ \pi^{*} $ generates such that $ \alpha $( $ \pi^{*} $)=$ \alpha $($ \pi $). 
In order to generate a permutation in P$_{n} $ from a permutation in P$ _{n-1} $ one can insert n$ - $1 in any of the n positions (n$ - $2 internal and two extreme positions).
$ \textbf{Case(i-a):} $
If $\pi^{*}_{n-1} $=n$ - $2 then by inserting n$ - $1 either we increase the number of
adjacencies by one if $\pi_{n}$=n$ - $1 or decrease the number of adjacencies by at least one.
That is, if n$-$1 is not placed in the last position then the existing adjacency of the last element of $ \pi^{*}_{n-1} $ i.e. n$ - $2 is automatically broken because after inserting n$ - $1, in P$_{n}$, n$ - $ 2 is not the largest element.
Further n$ - $1 can break an existing adjacency; thus for this sub-case $ \alpha $($ \pi $)$ \in $$\{$k+1, k$ - $1, k$ - $2$\}$; that is this sub-case is infeasible.
$ \textbf{Case(i-b):} $
If $ \pi^{*}_{n-1} $$ \neq $ n$ - $2 then if n$ - $1 is inserted in
a position where it does not create or break an adjacency then $ \alpha $($ \pi^{*} $)=$ \alpha $($ \pi $). There are n$ - $k$ - $2 such, positions, where n denotes the number of positions where n$ - $1 can be inserted, k of the excluded positions correspond to existing adjacencies in $ \pi^{*} $ and the remaining two excluded positions correspond to $ \pi_{n} $ and the position immediately following n$ - $2.
 Note that, f(n$ - $1, k)$-$f(n$ - $2, k$ - $1) denotes the number of permutations where $ \pi ^{*}_{n-1}$$ \neq $ n$ - $2 and $ \alpha $($ \pi ^{*}$)=k . Thus, the contribution of f(n$ - $1, k) to f(n, k) is
f(n$ - $1, k)$-$f(n$ - $2, k$-$1)*(n$ - $k$ - $2).\\
$ \textbf{Case(ii):}$ p= k$-$1. If $ \pi ^{*}_{n-1}$ =n$ - $2 then $ \pi_{n}$=n$ - $1 is the only possibility the corresponding contribution is f(i$ - $2, j$ - $2). If $ \pi ^{*}_{n-1}$$ \neq $ n$ - $2 then $ \pi_{n}$=n$ - $1 or n$ - $1 can be inserted immediately after n$ - $2. Thus, the
contribution of f(n$-$1, k) to f(n, k) is(f(n$ - $1, k$ - $1)$ - $f(n$ - $2, k$ - $2))*2+f(n$ - $2, k$ - $2).\\
$ \textbf{Case(iii):}$ p = k+1. If $ \pi ^{*}_{n-1}$ $ \neq $n$ - $2 any of the existing k+1 adjacencies can be broken. Otherwise, $ \pi_{n} $$ \neq $ n$ - $1 and n$ - $1 does not break any of the existing k adjacencies. Thus, the contribution of f(n-1, k+1) to f(n, k) is \\(f(n$ - $1, k+1)$ - $f(n$ - $2, k))*(k+1)+f(n$ - $2, k)*(n$ - $k$ - $1).\\
$ \textbf{Case(iv):} $ p=k+2. The only feasibility is that $ \pi ^{*}_{n-1}$ = n$ - $2 and n$ - $1 breaks one of the existing adjacencies. Thus, the contribution of f(n$ - $1, k+1) to f(n, k) is f(i$ - $ 2, j+1)*(j+1).
The theorem follows (the proof for Theorem \ref{th_type2_adj} is similar). The corresponding algorithm, $\textbf{Adjacency\_Count2} $, is given below.$~\blacksquare$\\
{\footnotesize
\begin{algorithmic}
\State Algorithm :$ \textbf{Adjacency\_Count2} $
\State Initialization: f[2][0]= 1; f[2][1]= 0; f[2][2]= 0; f[2][3]= 1; f[3][0]= 1; f[3][1]= 4; f[3][2]= 0; f[3][3]=0; f[3][4]=1;
\\

\For{(i=4,....n)}
\For{(j=0,....i)}
 \If{(j==0) } f[i][j] $ \gets $ (f[i$ - $1][j]) *(i$ - $2) + ( f[i$ - $1][j+1]$ - $f[i$ - $2][j])* (j+1) +\\$ \hspace{3.5cm} $ f[i$ - $2][j] *(i$ - $1$ - $j)+ f[i$ - $2][j+1] *(j+1);
 \Else 
   \If{(j==i) } f[i][j] $ \gets $( (f[i$ - $1] [j$ - $1])  $ - $  f[i$ - $2][j$ - $2]$ )* 2 + $ f[i$ - $2][j$ - $2] 
    \Else 
      \If{(j==1) } f[i][j] $ \gets $( (f[i$ - $1] [j$ - $1])*2  
          \Else~ f[i][j] $ \gets $( (f[i$ - $1] [j$ - $1])  $ - $  f[i$ - $2][j$ - $2]$ )* 2 + $ f[i$ - $2][j$ - $2] 
         \State  f[i][j] $ \gets $ f[i][j]+( (f[i$ - $1] [j+1])  $ - $  f[i$ - $2][j]$ )* (j+1) + $ f[i$ - $2][j ]*(i$ - $j$ - $1) ;
         \If{i$ - $j$ - $2 $ > $0} 
         f[i][j] $ \gets $ f[i][j]+( (f[i$ - $1] [j])  $ - $  f[i$ - $2][j$ - $1] )* (i$ - $j$ - $2)
         \State 
          f[i][j] $ \gets $ f[i][j]+f[i$ - $2] [j+1])*(j+1)    $ \setminus $$ \setminus $Break one adjacency from f[i$ - $2][j+1]
         \EndIf
      \EndIf    
    \EndIf
 \EndIf
\EndFor
 \EndFor
\end{algorithmic}

}
\end{document}